\documentclass[aps,nofootinbib,longbibliography,prd,twocolumn]{revtex4-1}
\usepackage[babel]{csquotes}
\usepackage{graphicx}
\usepackage{amsmath,amssymb}
\usepackage[colorlinks,citecolor=blue,linkcolor=blue,urlcolor=blue]{hyperref}
\usepackage{mathrsfs}
\usepackage{enumerate}
\def\be{\begin{equation}}
\def\ee{\end{equation}}

\begin{document}

\title{Planck stars: new sources in radio and gamma astronomy?\\
{\rm\normalsize\em Nature Astronomy 1 (2017) 0065}}
\author{Carlo Rovelli}
\affiliation{\small CPT, Aix-Marseille Universit\'e, Universit\'e de Toulon, CNRS, F-13288 Marseille, France.}

\maketitle

\begin{figure}[b]
\includegraphics[height=4.5cm]{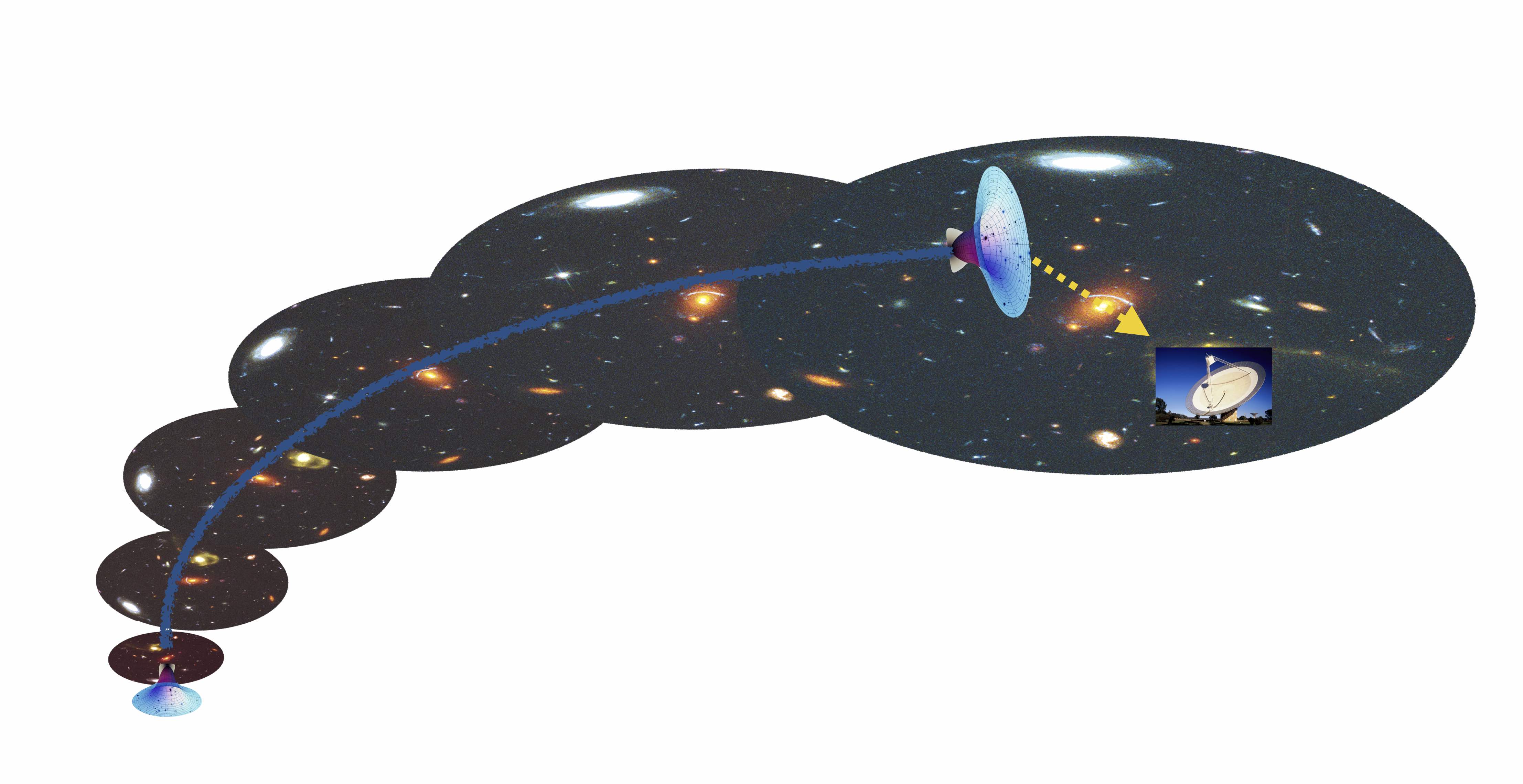}
\caption{Illustration of the Planck star phenomenology: matter collapses in the early universe forming a black hole that undergoes a rapid bounce. Because of the huge gravitational redshift, the subsequent explosion happens a cosmological external time later, producing signals we may observe.\vspace{1cm}}
\label{3}
\end{figure}

A new phenomenon, recently studied in theoretical physics, may have considerable interest for astronomers: the explosive decay of old primordial black holes via quantum tunnelling.  Models predict radio and gamma bursts with a characteristic frequency-distance relation making them identifiable.  Their detection would be of major theoretical importance. 
 
The expected signal may include two components \cite{Barrau2014c}: \emph{(i)} strong impulsive emission in the high-energy gamma spectrum ($\sim TeV$), and \emph{(ii)} strong impulsive signals in the radio, tantalisingly similar to the recently discovered and ``very perplexing"  \cite{Gibney2016} Fast Radio Bursts.   Both  the gamma and the radio components are expected to display a characteristic flattening of the cosmological wavelength-distance relation, which can make them identifiable \cite{Barrau2014b,Barrau2015}.  

The physics governing the decay is not exotic---in fact, it is conservative: just general relativity and quantum mechanics, physically reliable theories. However, lacking a consensual theory of quantum gravity, current models are hypothetical.  Detection and identification of these signals would represent the first direct observation of a quantum-gravitational phenomenon. 

A striking conclusion from the observations of the last decade is that our universe teems with black holes of widely different masses, spanning at least nine orders of magnitude---a conclusion reinforced by the recent gravitational-waves detection of the merger of two black holes of unexpected mass \cite{Abbott2016}. Black holes are stable according to classical general relativity, but there is theoretical consensus that they decay via quantum processes. Until recently, the only decay channel studied was Hawking evaporation \cite{Hawking}, a perturbative phenomenon too slow to have astrophysical interest: evaporation time of a stellar black hole is $10^{50}$ Hubble times.  

What can bring black hole decay within potential observable reach is a different, non-perturbative, quantum phenomenon: tunnelling, the same phenomenon that triggers nuclear decay in atoms. The explosion of a black hole out of its horizon is forbidden by the classical Einstein equations but classical equations are violated by quantum tunnelling.  Violation in a \emph{finite} spacetime region turns out to be sufficient for a black hole to tunnel into a white hole and explode \cite{Haggard2014}. The phenomenon does not violate causality, since it is the spacetime causal structure itself to tunnel. 

In the process, the gravitationally collapsing matter falls inside its horizon until its density reaches the Planck density, namely the quantity with the dimension of a density determined by the Planck constant $\hbar$, the Newton constant $G$ and the speed of light  $c$---these are the quantities that set the scale of quantum-gravitational phenomena.  At this stage, called ``Planck star" \cite{Rovelli2014}, the star can still be much larger than the Planck length, but quantum effects are expected to make gravity  strongly repulsive \cite{Ashtekar:2013hs}, triggering a bounce. The phenomenon is similar to the ``quantum pressure" that prevents electrons from falling into an atomic nucleus.  

Remarkably, because of the huge general-relativistic gravitational time dilation involved, collapse and bounce can be fast (milliseconds) in the proper time of the collapsing matter, and extremely slow (millions of years) in the external time. Thus the black holes we see in the sky can be bouncing Planck stars during their deep bounce phase, observed in extreme slow motion because of the very large gravitational time dilation. 

This phenomenon is plausible on the basis of our current understanding of gravity and quantum theory; the theoretical uncertainty regards the scale of the decay time.  If it is exponentially suppressed like generic macroscopic quantum tunnelling, it has no astrophysical consequences either. But the dumping exponential factor may be balanced by the phase-space factor due to the large black hole entropy, and arguments have been given  \cite{Haggard2014} indicating that the decay time could be of the order of 
\be
\tau\sim \frac{m^2}{m^2_{\scriptscriptstyle \rm Planck}}\ t_{\scriptscriptstyle  \rm Planck}
\ee
 where $m$ is the mass of the hole, and $t_{\scriptscriptstyle \rm Planck}$ and $m_{\scriptscriptstyle \rm Planck}$ are the Planck time and mass.  A detailed calculation of the hole decay time from first principles is undertaken in \cite{Christodoulou2016} using loop quantum gravity.  
 
For a black hole of planetary mass, equation (1) gives a lifetime of the order of the current Hubble time. This implies that primordial black holes --black holes formed by the large thermal fluctuations of the early universe-- of such mass may be exploding today.  Such an explosion should release an energy of the order 
 \be
 E=mc^2 \sim 10^{47} erg,
 \ee
or the mass of a small planet, exploding suddenly from a region of millimetre size.  See Figure \ref{3} for an artistic illustration of the phenomenon. 

\begin{figure}[t]
\includegraphics[height=5cm]{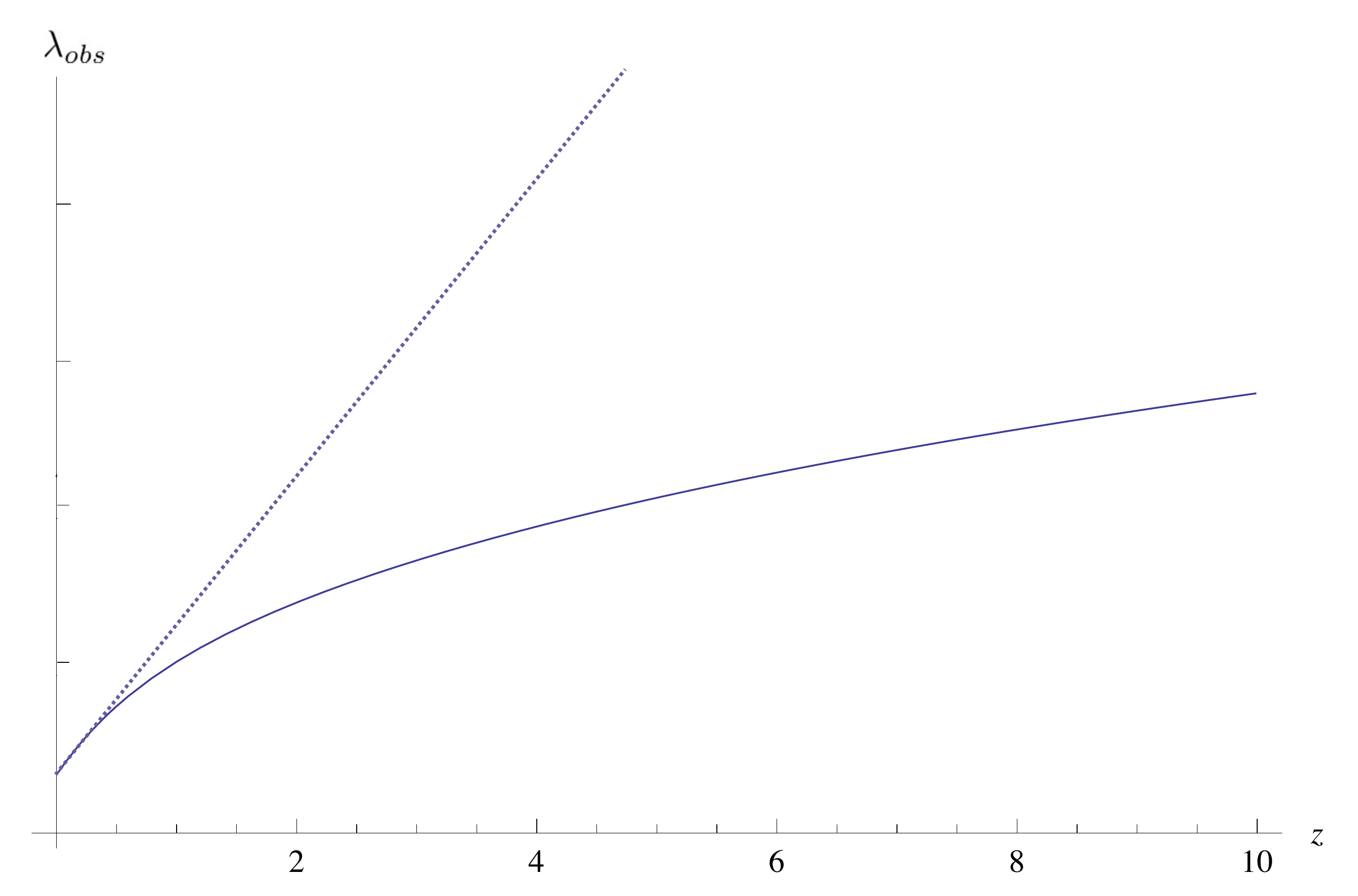}
\caption{The flattened wavelength-distance relation equation (3) (continuous line) for the radio component of the Planck star signal, compared with the strandard redshift (dotted).}
\label{1}
\end{figure}

\begin{figure}[t]
\includegraphics[height=5cm]{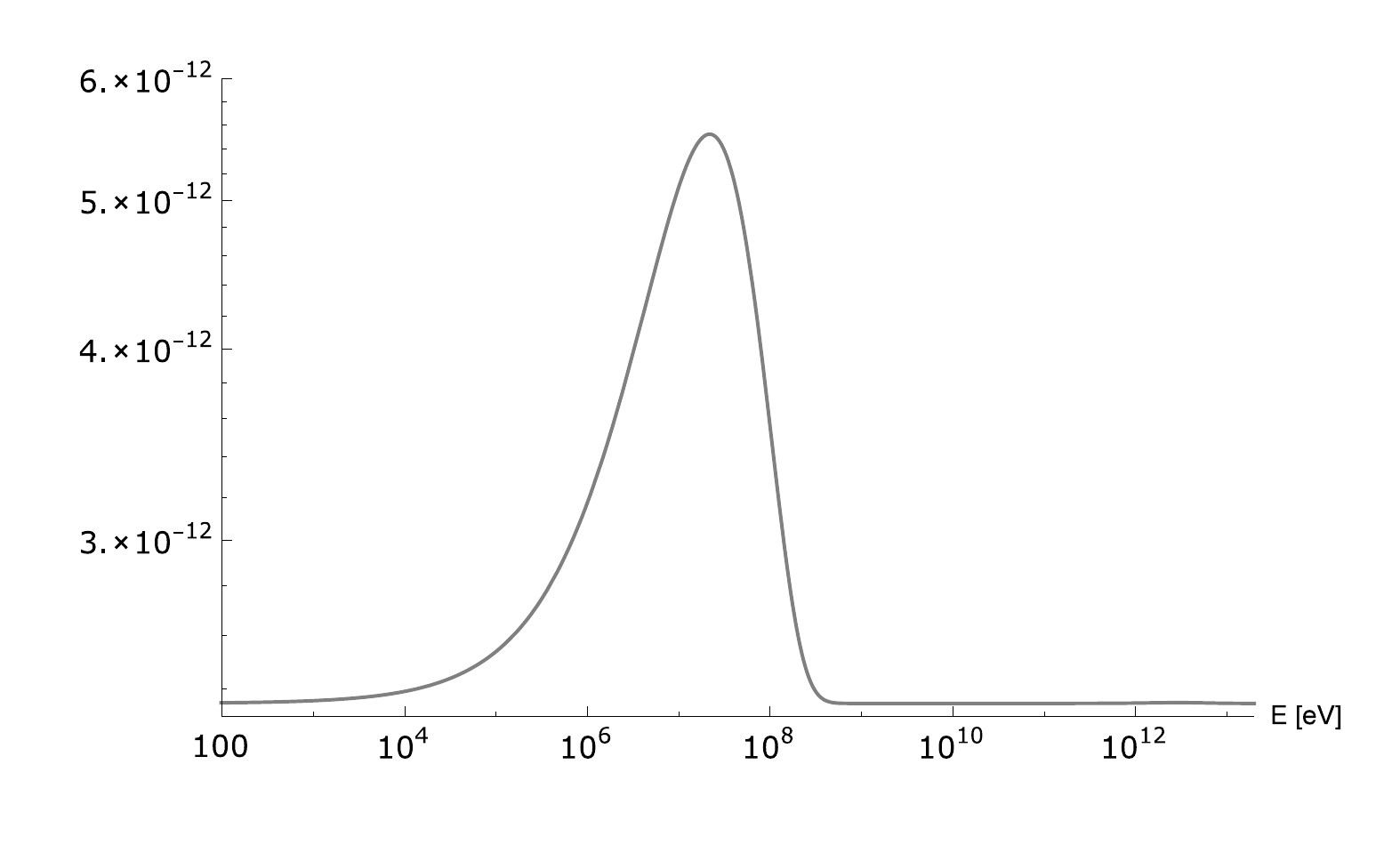}
\caption{Spectrum of the integrated emission of the gamma component of the Planck star signal   \cite{Barrau2015}. }
\label{2}
\end{figure}

A low-energy component of the signal emitted at the explosion should be a powerful burst with wavelength of the order of the size of the hole, thus around the millimetre range \cite{Barrau2014b}.  This is the predicted signal tantalisingly close to the observed Fast Radio Bursts.  

A second, high-energy, component is the photon gas originally collapsed in the early universe, liberated after the short-internal/long-external time. According to standard primordial black hole formation theory, the scale of the energy of these photons is determined by formation time, which in turn depends on the mass, and for millimetres black holes is expected in the $TeV$ range. 

The interesting aspect of the predicted signals is that their frequency-distance relation is expected to be different from the standard cosmological redshift. This is because holes we see exploding at cosmological distances have exploded in the past,  therefore had a shorter life, and therefore, according to equation (1), should be smaller. For the radio component, smaller mass implies smaller size and therefore shorter emitted wavelength.  For the gamma component, smaller mass implies, according to standard primordial black hole formation theory, a slightly earlier formation, when the plasma was hotter; this gives hotter photons trapped into the hole, and therefore, again, a shorter emission wavelength at explosion time (the internal proper time of the bounce is short, not allowing much internal evolution). Both emitted frequencies are thus higher for black holes we see exploding further away from us, partially compensating the cosmological redshift. The resulting flattened wavelength-distance relation is \cite{Barrau2014b}
\be
\small
\lambda_{obs}\sim \frac{2Gm}{c^2} (1+z)
  \sqrt{\frac{H_0^{-1}}{6\,k\Omega_\Lambda^{\,1/2}
}\ \sinh^{-1}\!\! \sqrt{\frac{\Omega_\Lambda}{\Omega_M(z+1)^3}}}
\ee
($z$ is the redshift factor, $H_0, \Omega_\Lambda, \Omega_M$ are the Hubble constant and the cosmological-constant and matter densities) and is depicted in Figure \ref{1}.  If observed, it would represent the smoking gun for identifying the Planck stars signals. 

This modified  redshift curve affects also the shape of the spectrum of the diffuse background due to the integrated emission of a population of bouncing black holes, opening up the possibility of revealing these signals as components of the cosmic ray  background \cite{Barrau2015}. The resulting spectrum looks like a slightly distorted (by the redshift/distance integration) blackbody, depicted in Figure \ref{2} for its high-energy component, a shape not expected from any other known astrophysical phenomenon.

Fast Radio Bursts observations are rapidly improving. CHIME is expected to observe dozens a day before the end of the year. 
Radio telescopes working in the radio, such as ALMA and SKA, could detect the low energy signal. Bursts are now receiving increasing attention, but the integrated emission may be easier to analyse.  Theoretical models favour signals with shorter wavelengths around $\lambda_{obs}\sim .2mm$.  There are detectors operating at these wavelengths, such as the Herschel instruments. The 200 micron range can be observed both by PACS and SPIRE.  The predicted signal falls in between PACS and SPIRE sensitivity zones. A problem is that the bolometer technology makes detecting short black-hole bursts difficult: they are likely to be mistaken for cosmic ray noise.  For the high energy component, the Fermi-LAT data could be particularly relevant.

Theoretical research in quantum gravity has been moving increasingly closer to phenomenology. The possibility of observing a quantum gravitational phenomenon is not anymore considered remote by theoreticians, and a number of possibilities have been suggested (see for instance \cite{giddings}). Because of the smallness of the Planck scale, the observation of a quantum gravitational phenomenon requires a large multiplicative factor. For a Planck star, the large ratio of the Hubble time $t_{\scriptscriptstyle \rm Hubble}$ to the Planck time $t_{\scriptscriptstyle  \rm Planck}$ provides such a factor, scaling up the Planck length $L_{\scriptscriptstyle  \rm Planck}$ to millimetres. Equation (1) gives indeed
 \be
\sqrt{\frac{t_{\scriptscriptstyle \rm Hubble}}{t_{\scriptscriptstyle  \rm Planck}}}\ L_{\scriptscriptstyle  \rm Planck}\sim 1\, mm. 
\ee
This is how a Planckian phenomenon can yield an effect at a macroscopic wavelength. 

Exploding black holes, or ``Planck stars", represent a speculative but realistic possibility to observe quantum gravity effects.  Detecting and identifying their signal in the sky would be of immense scientific value.

\end{document}